\documentclass[a4paper,UKenglish,cleveref, autoref, thm-restate]{lipics-v2021}

\usepackage{xcolor}
\usepackage{hyperref}
\usepackage{amsmath}

\hideLIPIcs
\nolinenumbers

\title               {A 4.542-approximation for the Small Cuts Cover problem via a multi-rate primal-dual algorithm}
\titlerunning{A 4.542-approximation for Small Cuts Cover via a multi-rate primal-dual algorithm}

\author{Zeev Nutov}{The Open University of Israel}{nutov@openu.ac.il}
{https://orcid.org/0000-0002-6629-3243}{}
\author{Elazar Koren}{The Open University of Israel}{elazarkoren@gmail.com}{}{}
\authorrunning{Zeev Nutov and Elazar Koren}
\Copyright{Zeev Nutov and Elazar Koren}


\begin{document}

\maketitle

\def\sem    {\setminus}
\def\subs   {\subseteq}
\def\empt   {\emptyset}

\def\wh   {\widehat}

\def\ga {\gamma}
\def\de   {\delta}

\def\bo   {\boldmath}
\def\RR  {{\cal R}}
\def\BB  {{\cal B}}

\def\CC  {{\cal C}}
\def\FF   {{\cal F}}
\def\LL   {{\cal L}}

\def\SS   {{\cal S}}
\def\TT   {{\cal T}}

\def\HH   {{\cal H}}
\def\MM  {{\cal M}}

\keywords{primal dual method, small cuts, pliable set-family, approximation algorithm}

\begin{abstract}
In the {\sc Small Cuts Cover} problem  we seek to cover by a min-cost edge-set 
the set-family of cuts of size $<k$ of a graph. 
Bansal, Cheriyan, Grout, and Ibrahimpur [ICALP 2023: 15:1--15:19] showed that 
the classic primal-dual algorithm with reverse delete 
of Williamson, Goemans, Mihail, and Vazirani [STOC 1993: 708--717]  
achieves approximation ratio $16$ for this problem. 
The analysis of the approximation ratio was then improved in a sequence of papers 
\cite{N-prel,B,N-MFCS}, culminating in the $5$-approximation \cite{SBC}.
A matching example \cite{N-tight} shows that this bound is tight for the classic algorithm.
We improve the approximation ratio to $(3+\sqrt{37})/2<4.542$ using 
a {\em multi-rate} primal-dual algorithm (that applies for strongly pliable set-families)
in which the dual variables of the 
inclusion-minimal violated sets are not raised uniformly, but with different rates. 
We also provide a tight example.
To the best of our knowledge, this is the first example in network design where for the standard LP, 
an efficiently computable choice of growth rates improves on the tight guarantee 
of uniform growth.
\end{abstract}

\section{Introduction} \label{s:intro}

For an edge set or a graph $J$ on node set $V$ and $S \subs V$ 
let $\de_J(S)$ denote the set of edges with exactly one end in $S$ 
and let $d_J(S)=|\de_J(S)|$ denote their number.
An edge set $J$ {\bf covers} $S$ if $d_J(S) \geq 1$.
We will consider the following problem, in which we need to cover by a min-cost edge-set 
the family of cuts of size $< k$ of a graph $H$.

\begin{center}
\fbox{\begin{minipage}{0.98\textwidth} \noindent
\underline{\sc Small Cuts Cover} \\ 
{\em Input:} A graph $H=(V,E_H)$, an edge set $E$ on $V$ with costs $\{c_e \ge 0:e \in E\}$, an integer $k$. \\ 
{\em Output:} A min-cost edge set $J \subs E$ that covers the family $\FF^k_H=\{\empt \ne S \subset V:d_H(S)<k\}$.
\end{minipage}} \end{center}

The following generic meta-problem captures dozens of specific  network design problems,
including {\sc Small Cuts Cover}.

\begin{center}
\fbox{\begin{minipage}{0.98\textwidth} \noindent
\underline{\sc Set-Family Edge-Cover} \\ 
{\em Input:} \ \ A graph $G=(V,E)$ with edge costs $\{c_e \ge 0:e \in E\}$, a set family $\FF$ on $V$. \\ 
{\em Output:} A min-cost edge set $J \subs E$ such that $d_J(S) \geq 1$ for all $S \in \FF$.
\end{minipage}} \end{center}

In this problem the family $\FF$ may not be given explicitly. We will assume that
$\empt,V \notin \FF$ (otherwise the problem has no feasible solution) and 
that some queries related to $\FF$ can be answered in time polynomial in $n=|V|$. 
An inclusion-minimal set in $\FF$ is called an $\FF$-{\bf core}, or just a {\bf core}, if $\FF$ is clear from the context. 
Following previous work, we will assume that for any edge set $I$, 
the cores of the {\bf residual family  $\FF^I=\{S \in \FF:d_I(S)=0\}$} of $\FF$ 
(the family of sets in $\FF$ that are uncovered by $I$) can be computed in time polynomial in $n=|V|$.
Such a procedure is available for the family $\FF_H^k$ of the {\sc Small Cuts Cover} problem (see \cite{BCGI}),
that also has the following properties.

\begin{definition}
Two sets $A,B$ {\bf overlap} if the three sets $A \cap B,A \sem B,B \sem A$ are non-empty;
if also $A \cup B \ne V$ then $A,B$ {\bf cross}.
A set family $\FF$ is {\bf pliable} if for any $A,B \in \FF$ at least two of
the sets $A \cap B,A \cup B,A \sem B,B \sem A$ are in $\FF$, and it is 
{\bf strongly-pliable} if in addition 
it satisfies the {\bf structural submodularity} property: 
whenever $A,B \in \FF$ cross
one of $A \cap B,A \cup B$ is in $\FF$ and 
one of $A \sem B,B \sem A$ is in $\FF$.
\end{definition}

If $\FF$ is strongly pliable (and $\empt, V \notin \FF$) then 
for any $A,B \in \FF$ the following holds:
\begin{eqnarray}
\mbox{If } A \cap B \ne \empt \mbox{ and } A \cup B \ne V                  & \mbox{then} & A \cap B \in \FF \mbox{ or } A \cup B \in \FF. \label{e:iu} \\
\mbox{If } A \sem B \ne \empt \mbox{ and } B \sem A \ne \empt \ \ & \mbox{then} & A \sem B \in \FF \mbox{ or } B \sem A \in \FF. \label{e:diff}
\end{eqnarray}
If $A,B$ cross this follows from structural submodularity.
Otherwise, this follows from pliability and the assumption $\empt, V \notin \FF$.
Note that conditions (1,2) alone do {\em not} imply strong pliability;
e.g., $\FF=\{A,B,A \sem B\}$ for overlapping $A,B$ with $A \cup B=V$.
However, if $\FF$ is {\em symmetric} --  $A \in \FF$ implies $V \sem A \in \FF$, 
then (1,2) do imply strong pliability.\footnote{ 
In fact, for symmetric $\FF$ only one of the conditions (1,2) will suffice, 
as the other follows from symmetry. 
On the other hand, one can see that if $\FF$ is strongly pliable
then so is the ``symmetrized family'' $\FF \cup \{V \sem A:A \in \FF\}$ of $\FF$.
Since covering $\FF$ is equivalent to covering 
its symmetrized family,
we can just consider symmetric families that satisfy (\ref{e:iu}).
We nevertheless retain the terminology of previous work,
as it is convenient and makes the assumptions behind the individual statements explicit.

}

Williamson, Goemans, Mihail, and Vazirani \cite{WGMV} 
designed a primal-dual algorithm with reverse delete, and showed that it achieves 
approximation ratio $2$ for {\em uncrossable} $\FF$, when 
$A,B \in \FF$ implies $A \cap B,A \cup B \in \FF$ or $A \sem B,B \sem A \in \FF$.
In a breakthrough paper, Bansal, Cheriyan, Grout, and Ibrahimpur \cite{BCGI} showed that 
this algorithm achieves approximation ratio $16$ for {\sc Small Cuts Cover},
while \cite{SBC2} shows that a standard iterative rounding 
approach does not give a good ratio.
The analysis of the approximation ratio was then improved in a sequence of papers 
\cite{N-prel,B,N-MFCS}, culminating in the $5$-approximation 
of Simmons, Bansal and Cheriyan  \cite{SBC}.
A matching example of Nutov \cite{N-tight} shows that this bound 
is tight for the classic primal-dual algorithm.
Hence one cannot break the ratio $5$ just by a sharper analysis, 
but needs to change the algorithm.
We design a new algorithm and prove the following. 

\begin{theorem} \label{t:main}
The {\sc Set Family Edge Cover} problem with strongly-pliable set family $\FF$,
and thus also the {\sc Small Cuts Cover} problem,
admits approximation ratio $(3+\sqrt{37})/2 < 4.542$.
\end{theorem}

Interestingly, we beat the classic primal-dual algorithm using a generalization of it.
In the classic primal-dual algorithm \cite{WGMV,GW},
the dual variables of ``active sets'' are raised uniformly.
The active sets are chosen as the cores of the residual family, 
and most literature identifies the purely algorithmic term ``active sets'' 
with the combinatorial term ``cores''. 
A natural justification for the uniform rate is that the algorithm 
cannot foresee the final structure of the computed solution.
For strongly-pliable set families (and the {\sc Small Cuts Cover} problem), 
this traditional approach gives a tight approximation ratio of $5$. 
We challenge this long-standing convention by using a {\em multi-rate} primal-dual algorithm,
in which the dual variables of the cores are raised at different rates -- 
each core $C$ is raised at rate $r_C$. 
Our key insight is that certain cores can be classified as ``dangerous'';
by decelerating their dual growth we guide the algorithm to pick better edges. 
This multi-rate approach breaks the $5$-approximation barrier,
improving the approximation ratio to $(3+\sqrt{37})/2 < 4.542$.
To the best of our knowledge, 
this gives the first example of a fundamental network design problem 
where a multi-rate strategy with a standard LP strictly outperforms the classic uniform method.
It suggests that differential growth rates may be a powerful, yet underutilized, 
tool for obtaining better approximation guarantees for some network design problems. 

A multi-rate strategy was already considered as a theoretical possibility in the 
primal-dual schema of Borodin, Cashman and Magen \cite{BCM}. 
They note that ``... in some sense our algorithms may be as
strong as the recipe of choice of rates. 
If there is no limitation on this recipe, optimality is always achievable.''
Hence while the multi-rate strategy idea is not new, 
the main challenge was to turn it into an effective algorithm 
by choosing the rates efficiently and usefully.
Our rate rule is determined by a hierarchical structure of cores of the current residual family. 
For a core $C_j$ let $\wh C_j$ be the maximal set in the residual family 
that contains $C_j$ and no other core.
Put an arc $C_i \rightarrow C_j$  when a distinct core $C_i$ intersects $\wh C_j$; 
we say that then {\em $C_i$ hits $C_j$}.
We prove that the obtained hitting digraph $\HH$ is a forest of out-arborescences.
We partition the cores into blue and red cores, and find a matching $\MM$
in $\HH$ between blue cores. Let $\BB_0$ be the unmatched blue cores.
Every red core hits at least two cores in $\BB_0$, and every non-root
core in $\BB_0$ is hit by a red core.
The red cores are the ``dangerous'' ones -- they are raised at rate $r<1$,
while the blue cores are raised at rate $1$.
An approximation ratio $x$ balances two critical requirements:
$xr \ge 3+r$ and $x \ge 2+3r$.
The resulting values are $r=(\sqrt{37}-1)/6$ and $x=(3+\sqrt{37})/2$.
Thus the improvement comes from an efficiently computable structural choice of rates,
not from foreknowledge of the returned solution.

To prove the approximation ratio we use {\em hollow subtrees} -- 
a new generic tool to analyze the performance of primal-dual algorithms
when for any inclusion-minimal solution $I$ there exists a laminar {\em witness family} -- 
a family $\{A_e:e \in I\} \subs \FF$ such that $\de_I(A_e)=\{e\}$ for all $e \in I$.
A laminar family can be represented by a rooted tree $\TT$.
A set $A \in \TT$ {\em owns} a core $C$ if $A$ is the inclusion-minimal set in $\TT$
that contains $C$.
We call a rooted subtree $\TT'$ of $\TT$ {\em hollow}
if only its root and leaves may own cores.
To prove approximation ratio $x$ 
we assign $x \cdot r_C$ tokens to each core $C$, 
and show that the tokens owned by any maximal hollow subtree $\TT'$ can be redistributed 
such that edges and the root of $\TT'$ get enough tokens.

This general technique can be used to get relatively simple direct proofs 
of known results for the classic primal-dual algorithm.
For example, it easily implies 
ratio $2$ when no core crosses a set in $\TT$ (e.g., for uncrossable families \cite{WGMV}), 
as well as a simpler direct proof of ratio $5$ for strongly-pliable families \cite{SBC}. 
Moreover, it allows us to prove ratio $<5$ for our multi-rate primal-dual algorithm
by relatively simple local arguments. 
Note that the witness tree $\TT$ organizes token payments, 
while the hitting forest $\HH$ is used to select rates. 

Our paper only makes the first step in exploiting nonuniform dual growth
and we do not exclude that a finer treatment will yield further improvements. 
But beyond the specific approximation factor, 
we show how a structural property of the input set-family 
can guide the design of growth rates that improve on a tight uniform-growth guarantee.
Summarizing, our main contributions to design and analysis of primal-dual algorithms~are: 
\begin{enumerate}[(i)]
\item
Establishing and using a structural property of cores to choose dual growth rates. 
\item
An example of a cut-covering network design problem for which
a multi-rate strategy outperforms the uniform one, w.r.t. standard LP.
\item
A new method of analyzing the algorithm's performance via hollow subtrees. 
\end{enumerate}

{\sc Small Cuts Cover} and pliable families 
play an important role in designing approximation algorithms for 
 {\sc Capacitated $k$-Connected Subgraph} and 
{\sc Flexible Graph Connectivity} ({\sc FGC}), cf. \cite{AHM,BCGI, B,BCKS}.
For example, Bansal, Cheriyan, Khanna and Simmons \cite{BCKS}
show that an LP $\rho$-approximation for {\sc Small Cuts Cover} implies 
a $(\rho+2)$-approximation for {\sc $(1,q)$-FGC}, thus obtaining 
ratio $7$ for {\sc $(1,q)$-FGC}; our work reduces it to $6.542$. 

\medskip

This paper is organized as follows. 
In Sect.~\ref{s:prd} we describe the multi-rate primal-dual algorithm.
In Sect.~\ref{s:properties} we give properties of strongly pliable families and 
use them in Sect.~\ref{s:rs} to describe our rate system. 
In Sect.~\ref{s:overview} we prove ratio $\sqrt{13}+1<4.606$ based on a
structural lemma proved in Sect.~\ref{s:lemma}.
Sect.~\ref{s:improved-complete} gives 
the final ratio $(3+\sqrt{37})/2$ and a tight example.

\section{Multi-rate primal-dual algorithm} \label{s:prd}

We start by describing the multi-rate primal-dual algorithm for an arbitrary set-family $\FF$. 
Consider the following LP-relaxation {\bf (P)} for the {\sc Set Family Edge Cover} problem 
and its dual program {\bf (D)}:
\[ \displaystyle
\begin{array} {lllllll} 
& \hphantom{\bf (P)} & \min                 & \ \ \displaystyle \sum_{e \in E} c_e x_e                                                                 & 
    \hphantom{\bf (P)} & \max                & \ \ \displaystyle \sum_{S \in \FF} y_S  \\
& \mbox{\bf (P)}          & \ \mbox{s.t.} & \displaystyle \sum_{e \in \de_E(S)} x_e  \geq 1   \ \ \ \forall S \in \FF \ \ \ \ \ \ \  &
    \mbox{\bf (D)}          & \ \mbox{s.t.} & \displaystyle \sum_{\de_E(S) \ni e} y_S \leq c_e \ \ \ \forall e \in E \\
& \hphantom{\bf (P)} &                          & \ \ x_e \geq 0 \ \ \ \ \ \ \ \ \ \ \ \forall e \in E                                                                  &
    \hphantom{\bf (P)} &                          & \ \ y_S \geq 0 \hspace*{1.35cm} \forall S \in \FF 
\end{array}
\]

Given a solution $y$ to {\bf (D)}, an edge $e \in E$ is {\bf tight} if the inequality of $e$ in {\bf (D)} holds with equality.
The algorithm has two phases.

{\bf Phase~1} starts with $J=\empt$ and applies a sequence of iterations.
At the beginning of the iteration we assign a rate $r_C$ for every core $C$ of the residual family $\FF^J$.  
Then we raise the dual variables of every core $C$ by $\epsilon \cdot r_C$, 
until some edge $e \in E \sem J$ becomes tight, 
and add all tight edges in $E \sem J$ to $J$ in some fixed order.
Phase~1 terminates when $J$ covers $\FF$.

{\bf Phase~2} is a ``reverse delete'' phase, in which we process edges in the reverse order
that they were added, and delete an edge $e$ from $J$ if $J \sem \{e\}$ still covers $\FF$.
At the end of the algorithm, $J$ is output.

The produced dual solution is feasible, hence 
$\sum_{S \in \FF} y_S \leq {\sf opt}$, by the Weak Duality Theorem.
To prove an approximation ratio of $\rho$, it is sufficient  
to prove that at the end of the algorithm the following holds for the returned solution $J$
and the dual solution $y$:
\[
\sum_{e \in J} c(e) \leq \rho \sum_{S \in \FF} y_S \ .
\]
As any edge in the solution $J$ returned by the algorithm is tight, this is equivalent to
\[
\displaystyle \sum_{e \in J} \sum_{\de_J(S) \ni e} y_S \leq \rho \sum_{S \in \FF} y_S \ .
\]
By changing the order of summation we get:
\[
\sum_{S \in \FF} d_J(S) y_S \leq \rho \sum_{S \in \FF} y_S \ .
\]
It is sufficient to prove that at any iteration the increase at the left hand side is at most
the increase at the right hand side. 
Let us fix some iteration with the current partial solution $J$, 
let $\CC$ be the family of $\FF^J$-cores at this iteration,
and let $I$ be the set of edges picked after $J$ that survived the reverse-delete phase.
\begin{itemize}
\item
The increase in the left hand side is 
$ \displaystyle \epsilon \cdot \sum_{C \in \CC} r_C d_I(C)$.
\item
The increase in the right hand side is 
$ \displaystyle \epsilon \cdot \rho \sum_{C \in \CC} r_C$.
\end{itemize}
Since the edges were deleted in reverse order,
the edges in $I$ were considered for deletion when all edges in $J$ were still present.
Thus $I$ is an inclusion-minimal cover of $\FF^J$.
Consequently, we have the following.

\begin{corollary} \label{c:2C}
The multi-rate primal-dual algorithm achieves approximation ratio $\rho$ 
if for any edge set $J$ the following holds:
If $\{r_C:C \in \CC\}$ is the rate system for the family $\CC$ of $\FF^J$-cores 
and $I$ is an inclusion-minimal cover of $\FF^J$ then
\[
\sum_{C \in \CC} r_C d_I(C)  \leq  \rho \sum_{C \in \CC} r_C \ .
\]
\end{corollary}

\section{Some properties of strongly pliable families} \label{s:properties}

Let $\FF$ be a set-family. 
One can see that if an edge covers one of the sets $A \cap B,A \cup B,A \sem B, B \sem A$ 
then it also covers one of $A,B$.
This implies the following.

\begin{lemma} \label{l:res}
Let $\FF^I$ be a residual family of $\FF$.
If $A,B \in \FF^I$ then $A,B \in \FF$ and 
any of the sets $A \cap B, A \cup B, A \sem B, B \sem A$ that is in $\FF$ is also in $\FF^I$.
Consequently, if $\FF$ is pliable or has the structural submodularity property, then so is $\FF^I$. 
\end{lemma}

\begin{lemma} \label{l:disjoint}
Let $\FF$ be pliable and let $A \in \FF$ and $C \in \CC_\FF$.
If $A \cap C, C \sem A \ne \empt$ then $A,C$ cross and:
$A \cup C,A \sem C \in \FF$ and $A \cap C,C \sem A \notin \FF$.
Thus the $\FF$ cores are pairwise disjoint.
\end{lemma}
\begin{proof}
We cannot have $A \cap C \in \FF$ or $C \sem A \in \FF$ by the minimality of $C$.
Hence $A \cup C,A \sem C \in \FF$, since $\FF$ is pliable. 
Since $\empt,V \notin \FF$, $A,C$ must cross.  
If $C_1,C_2$ are two distinct cores and $C_1 \cap C_2 \ne \empt$, 
then $C_1 \sem C_2 \in \FF$, contradicting the minimality of $C_1$. 
\end{proof}

\begin{definition} [\cite{B}]
A set family $\FF$ has the {\bf Sparse Crossing Property} if any $A \in \FF$ crosses at most one $\FF$-core. 
\end{definition}

\begin{lemma} [\cite{SBC}] \label{l:sparse} 
If $\FF$ is strongly pliable then $\FF$ has the Sparse Crossing Property. 
\end{lemma}
\begin{proof}
Suppose to the contrary that two distinct $\FF$-cores $C_1$ and $C_2$ cross $A \in \FF$.
Then $A \cup C_1,A \cup C_2 \in \FF$ by Lemma~\ref{l:disjoint}.
Since $C_1,C_2$ are disjoint, neither $A \cup C_1$ nor $A \cup C_2$ contains the other. 
Thus by (\ref{e:diff}) at least one of the sets 
$(A \cup C_1) \sem (A \cup C_2)=C_1 \sem A$ and $(A \cup C_2) \sem (A \cup C_1)=C_2 \sem A$ 
is in $\FF$. This contradicts the minimality of $C_1,C_2$.
\end{proof}

\begin{definition} [\cite{SBC}]
A set family $\FF$ has {\bf \boldmath Property $\ga^*$} if for any $A \in \FF$ that crosses a core $C$
and any family $\{B_1, \ldots,B_p\} \subs \FF$ of pairwise-disjoint subsets of $A$ that all cross $C$, 
if the set $D=A \sem (B_1 \cup \cdots \cup B_p \cup C)$ is non-empty then it belongs to $\FF$. 
\end{definition}

\begin{lemma} [\cite{SBC}] \label{l:ga*}
If $\FF$ is strongly pliable then $\FF$ has Property $\ga^*$.
\end{lemma}
\begin{proof}
We prove by induction on $p \ge 1$ that if $D \ne \empt$ then $D \in \FF$.
The induction base case is $p=1$.
Then $B_1 \cup C \in \FF$ by Lemma~\ref{l:disjoint}.
None of the sets $B_1 \cup C,A$ contains the other, hence by (\ref{e:diff})
$(B_1 \cup C) \sem A \in \FF$ or $A \sem (B_1 \cup C) \in \FF$.
Since $(B_1 \cup C) \sem A=C \sem A \subset C$, the latter must hold.

For the induction step we will assume that the lemma holds for $B_1, \ldots, B_{p-1}$ 
and prove it for $B_1, \ldots, B_p$.
Let $D'=A \sem (B_1 \cup \cdots \cup B_{p-1} \cup C)$.
Note that:
\begin{itemize}
\item
$B_p \sem D'=B_p \cap C \ne \empt$ since $B_p,C$ cross. 
\item
$D=D' \sem B_p \ne \empt$ by the assumption.
\end{itemize}
Hence by (\ref{e:diff}),
$B_p \sem D' \in \FF$ or $D=D' \sem B_p \in \FF$.
Since $B_p \sem D'=B_p \cap C \subset C$, 
the latter must hold, concluding the proof.
\end{proof}

\begin{definition}
A set-family $\LL$ is {\bf laminar} if no two sets in $\LL$ overlap, namely, 
if any two sets in $\LL$ are disjoint or one of them contains the other. 
Given an inclusion-minimal cover $I$ of $\FF$ 
we say that a set-family $\LL \subs \FF$ is a {\bf witness family} for $I$ if 
for any $e \in I$ there is a unique set $A_e \in \LL$ such that $\de_I(A_e)=\{e\}$
and $\LL=\{A_e:e \in I\}$.
\end{definition}

Note that if $I$ is an inclusion-minimal cover of $\FF$ then there exists a witness family 
$\LL \subs \FF$ for $I$.
The next lemma states that there always exists such $\LL$ that is laminar.

\begin{lemma} [\cite{BCGI}] \label{l:lami}
For any inclusion-minimal edge-cover of a pliable set family $\FF$ there exists 
a witness family $\LL$ that is laminar.
\end{lemma}
\begin{proof}
Write $\de(S)=\de_I(S)$.
Let $\LL\subs\FF$ be a laminar family of witnesses for distinct edges of $I$,
initially empty. By minimality of $I$, an unrepresented edge $f$ has a witness $B$.
We show how to add a witness for $f$, possibly changing members of $\LL$.

Let $\LL_B$ consist of the members of $\LL$ overlapping $B$.
If $\LL_B=\empt$, add $B$.
Otherwise, both ends of $f$ lie on the same side of each member of $\LL_B$.
If some member contains neither end, choose an inclusion-minimal such member $A$.
Otherwise $\LL_B$ is a chain, and we choose its largest member $A$.

Let $\de(A)=\{e\}$. The sets $A\cap B,A\cup B,A\sem B,B\sem A$
have cuts $\empt,\{e\},\{f\},\{e,f\}$ in some order:
the only edges between the regions determined by $A,B$ are $e,f$,
which form a two-edge path after contracting the non-empty regions.
Since $I$ covers $\FF$, the zero-cut set is not in $\FF$.
Thus pliability gives among these sets a witness $W\in\FF$ for $e$ or $f$.
Each of the four sets is laminar with $A,B$ and with every set laminar with both.

If $W$ witnesses $f$, replace $B$ by $W$.
Otherwise replace $A$ by $W$. This preserves laminarity:
in the first case of the choice of $A$, $W=A\cap B$ or $A\sem B$,
and no member of $\LL_B$ is properly contained in $A$;
in the second case, $W=A\cup B$ or $B\sem A$,
and every other member of $\LL_B$ is contained in $A$.
Members outside $\LL_B$ are laminar with both $A,B$, hence with $W$.
In either replacement the number of members of $\LL$ overlapping the witness for $f$ decreases.
Repeating, we can add a witness for $f$, and eventually represent every edge of $I$.
\end{proof}

\section{Hierarchical property of cores and the rate system} \label{s:rs}

We partition the cores into blue and red cores, raised at rates $1$ and $r<1$, respectively. 
We need some definitions and lemmas to describe this partition.

\begin{definition}
The {\bf halo family} $\FF(\CC)$ of a subfamily $\CC \subs \CC_\FF$ of $\FF$-cores is 
the family of sets in $\FF$ that contain every core $C \in \CC$ 
but contain no other core, namely,
\[
\FF(\CC)=\{S \in \FF: \{C \in \CC_\FF:C \subs S\}=\CC\} \ .
\]
\end{definition}

A set-family is a {\bf nested family} 
if $A \subset B$ or $B \subset A$ for any two sets $A,B$ in the family.

\begin{lemma} \label{l:FC}
If $\FF$ is strongly pliable then $\FF(\CC)$ is a nested family for any $\CC \subs \CC_\FF$.
\end{lemma}
\begin{proof}
Suppose to the contrary that there are $A,B \in \FF(\CC)$ that overlap.
By (\ref{e:diff}), one of $A \sem B, B \sem A$ is in $\FF$, say $A \sem B \in \FF$. 
This implies that $A \sem B$ contains an $\FF$-core $C \in \CC_\FF \sem \CC$, 
contradicting that $A \in \FF(\CC)$.
\end{proof}

\begin{definition} \label{d:H}
We say that a core $C_i$ {\bf hits} a core $C_j$ for $i \ne j$ and write $C_i \rightarrow C_j$ 
if $C_i \cap \wh C_j \ne \empt$, where $\wh C_j$ denotes the maximal set in $\FF(C_j)$.
The {\bf hitting digraph} $\HH$ of $\FF$ has node set $\CC_\FF$ 
and there is an arc $C_i \rightarrow C_j$ if $C_i$ hits $C_j$. 
\end{definition}

We will prove the following.

\begin{lemma} \label{l:arb}
If $\FF$ is strongly pliable then the hitting digraph $\HH$ 
is a collection of node-disjoint out-arborescences. 
\end{lemma}
\begin{proof}
$\HH$ has maximum indegree $1$ by the Sparse Crossing Property.
We prove that $\HH$ has no directed cycles.
Let $C_1 \rightarrow C_2 \rightarrow \cdots \rightarrow C_p$
be a directed simple path in $\HH$ 
and let $\wh C_i$ be the maximal set in $\FF(C_i)$.
Let $\CC_p=\{C_1, \ldots,C_p\}$ and $S_p=\wh C_1 \cup \cdots \cup \wh C_p$.
We claim that
\begin{equation} \label{e:path}
\mbox{If $S_p \ne V$ then $S_p \in \FF(\CC_p)$.} 
\end{equation}
Assume $S_p\ne V$. We first prove that $S_p \in \FF$ by induction on $p$, see Fig.~\ref{f:arb}. 
The induction base case $p=1$ is trivial. 
Suppose that $p \ge 2$. By the induction hypothesis if $S_{p-1} \ne V$ then $S_{p-1} \in \FF$.

Note that $S_{p-1} \cap \wh C_p$ contains no core;
this is since the only core contained in $\wh C_p$ is $C_p$, 
but $C_p \not\subs S_{p-1}$ since 
$C_p$ is disjoint from $\wh C_2,\ldots,\wh C_{p-1}$ and $C_p \not\subs \wh C_1$.
Note that $S_{p-1} \cap \wh C_p$ contains a node from $C_{p-1}$.
Thus by (\ref{e:iu}) $S_{p-1} \cap \wh C_p \in \FF$ or $S_{p-1} \cup \wh C_p \in \FF$.
Since $S_{p-1} \cap \wh C_p$ contains no core, $S_p=S_{p-1} \cup \wh C_p \in \FF$, as required.

\begin{figure} \centering

\begingroup
\setlength{\unitlength}{0.375bp}
\setlength{\fboxsep}{0pt}
\begin{picture}(859,390)
\put(0,0){\includegraphics[scale=0.5]{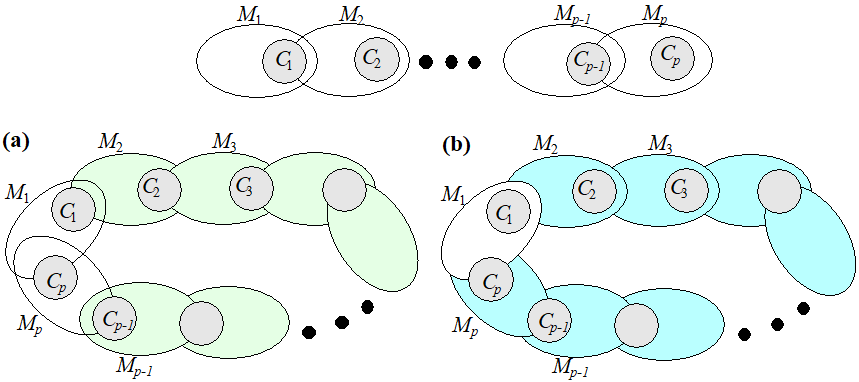}}
\put(236,358){\color{white}\rule{30\unitlength}{21\unitlength}}
\put(338,358){\color{white}\rule{28\unitlength}{21\unitlength}}
\put(551,357){\color{white}\rule{47\unitlength}{23\unitlength}}
\put(641,358){\color{white}\rule{31\unitlength}{21\unitlength}}
\put(5,178){\color{white}\rule{28\unitlength}{22\unitlength}}
\put(97,228){\color{white}\rule{29\unitlength}{22\unitlength}}
\put(210,229){\color{white}\rule{30\unitlength}{22\unitlength}}
\put(15,47){\color{white}\rule{28\unitlength}{22\unitlength}}
\put(114,4){\color{white}\rule{47\unitlength}{23\unitlength}}
\put(440,176){\color{white}\rule{29\unitlength}{22\unitlength}}
\put(531,227){\color{white}\rule{30\unitlength}{21\unitlength}}
\put(645,228){\color{white}\rule{29\unitlength}{22\unitlength}}
\put(452,45){\color{white}\rule{30\unitlength}{22\unitlength}}
\put(550,3){\color{white}\rule{47\unitlength}{23\unitlength}}
\put(249,369){\makebox(0,0){\colorbox{white}{\fontsize{6}{7}\selectfont$\wh C_1$}}}
\put(352,369){\makebox(0,0){\colorbox{white}{\fontsize{6}{7}\selectfont$\wh C_2$}}}
\put(574,369){\makebox(0,0){\colorbox{white}{\fontsize{6}{7}\selectfont$\wh C_{p-1}$}}}
\put(658,369){\makebox(0,0){\colorbox{white}{\fontsize{6}{7}\selectfont$\wh C_p$}}}
\put(17,189){\makebox(0,0){\colorbox{white}{\fontsize{6}{7}\selectfont$\wh C_1$}}}
\put(110,239){\makebox(0,0){\colorbox{white}{\fontsize{6}{7}\selectfont$\wh C_2$}}}
\put(224,239){\makebox(0,0){\colorbox{white}{\fontsize{6}{7}\selectfont$\wh C_3$}}}
\put(30,58){\makebox(0,0){\colorbox{white}{\fontsize{6}{7}\selectfont$\wh C_p$}}}
\put(136,15){\makebox(0,0){\colorbox{white}{\fontsize{6}{7}\selectfont$\wh C_{p-1}$}}}
\put(453,188){\makebox(0,0){\colorbox{white}{\fontsize{6}{7}\selectfont$\wh C_1$}}}
\put(546,239){\makebox(0,0){\colorbox{white}{\fontsize{6}{7}\selectfont$\wh C_2$}}}
\put(659,237){\makebox(0,0){\colorbox{white}{\fontsize{6}{7}\selectfont$\wh C_3$}}}
\put(466,56){\makebox(0,0){\colorbox{white}{\fontsize{6}{7}\selectfont$\wh C_p$}}}
\put(573,13){\makebox(0,0){\colorbox{white}{\fontsize{6}{7}\selectfont$\wh C_{p-1}$}}}
\end{picture}
\endgroup
\caption{
The set $S_p$ and the sets $A,B$ in the proof of Lemma~\ref{l:arb}.}
\label{f:arb} \end{figure}

We show that $S_p \in \FF(\CC_p)$.
Clearly, $C_i \subs S_p$ for $i=1, \ldots,p$.
Suppose to the contrary that $C \subs S_p$ 
for some $C \in \CC_\FF \sem \CC_p$. 
Note that $C$ is disjoint from each one of the sets $\wh C_2, \ldots \wh C_p$, 
since $\FF$ is sparse. Thus $C \subs \wh C_1$.
Since $C \ne C_1$, this contradicts that $\wh C_1 \in \FF(C_1)$.

Now suppose to the contrary that $\HH$ has a directed cycle
$C_1 \rightarrow C_2 \rightarrow \cdots \rightarrow C_p \rightarrow C_1$.
Let $\wh C_i$ be the maximal set in $\FF(C_i)$. 
We cannot have $p=2$ since then by (\ref{e:diff})
one of the sets $\wh C_1 \sem \wh C_2, \wh C_2 \sem \wh C_1$ is in $\FF$, 
but neither contains an $\FF$-core.  
Suppose that $p \ge 3$.
Let $\CC=\{C_2, \ldots,C_{p-1}\}$ and $B'= \wh C_2 \cup \cdots \cup \wh C_p$.
Consider two sets (see Fig.~\ref{f:arb}):
\begin{eqnarray*}
A & = & \wh C_2 \cup \cdots \cup \wh C_{p-1} \\
B & = & (\wh C_2 \cup \cdots \cup \wh C_p) \sem \wh C_1=B' \sem \wh C_1 
\end{eqnarray*}
By (\ref{e:path}) $A \in \FF(\CC)$ and $B' \in \FF$.  
We claim that $B=B' \sem \wh C_1 \in \FF(\CC)$. 
To see this, note that:
\begin{itemize}
\item
$\wh C_1 \sem B' \notin \FF$, since $\wh C_1 \sem B'$ is nonempty but contains no core; 
thus $B' \sem \wh C_1 \in \FF$, by (\ref{e:diff}).
\item
$C_2,\ldots,C_{p-1}$ are disjoint from $\wh C_1$, so they remain contained in $B$.
\item
$C_1 \not\subs B$ because $C_1\subs \wh C_1$ and 
$C_p \not\subs B$ since $C_p\cap \wh C_1\neq\empt$.
\item
No core outside the cycle is contained in $B'$, by (\ref{e:path}).
\end{itemize}

Now note that neither $A$ nor $B$ contains the other, since:
\begin{itemize}
\item
$A \cap C_1 \ne \empt$ while $B \cap C_1=\empt$;
\item
$A \cap C_p =\empt$ while $B \cap C_p \ne \empt$.
\end{itemize}
Since $A,B \in \FF(\CC)$, this contradicts Lemma~\ref{l:FC}. 
\end{proof}

\begin{lemma}
Let $\FF$ be a strongly pliable family and suppose that for any edge set (not necessarily a subset of $E$),
the cores of the residual family of $\FF$ can be computed in polynomial time.
Then for any core $C \in \CC_{\FF}$ the halo family $\FF(C)$ of $C$ can be computed in polynomial time.
Consequently, the hitting digraph $\HH$ of $\FF$ can be constructed in polynomial time.
\end{lemma}
\begin{proof}
Given a core $C$ we compute the halo family of $C$ as follows. 
Fix some node $s \in C$ and for every edge $sv$ with $v \in V \sem C$ check 
whether there exists an $\FF^{\{sv\}}$-core $A_v$ with $C \subset A_v$
that contains no $\FF$-core distinct from $C$.
The halo-family $\FF(C)$ of $C$ is the collection of the sets $A_v$ plus $C$.
Any $A_v$ chosen belongs to $\FF(C)$, and any $A \in \FF(C) \sem \{C\}$
is recovered as a set $A_v$ for $v \in A \sem B$, where $B$ is the 
maximal set in $\FF(C)$ properly contained in $A$. 
\end{proof}

\begin{lemma} \label{l:hits}
For any out-arborescence $\HH$ one can find in linear time a partition
$\BB,\RR$ of its nodes and a matching $\MM$ on $\BB$ in $\HH$ such that
if $\BB_0$ is the set of unmatched nodes in $\BB$ then:
\begin{enumerate}[(i)]
\item
Every node in $\RR$ has at least two children in $\BB_0$.
\item
Every non-root node in $\BB_0$ has its parent in $\RR$.
\end{enumerate}
\end{lemma}
\begin{proof}
Process the nodes bottom-up. A child is available if it is blue and unmatched.
A node with at least two available children is colored red.
Otherwise it is colored blue, and if it has one available child it is matched to this child.
A blue node left unmatched is thus either the root or has a red parent.
\end{proof}

\noindent
{\bf The rate system:}
For a partial solution $J$ compute the partition $\BB,\RR$ and matching $\MM$
of $\CC_{\FF^J}$ as in Lemma~\ref{l:hits}.
Let $r_C=1$ for $C\in\BB$ and $r_C=r$ for $C\in\RR$.
We first use $r=(\sqrt{13}-1)/3$; the improved choice
$r=(\sqrt{37}-1)/6$ is analyzed in Sect.~\ref{s:improved-complete}.

\section{Approximation ratio proof} \label{s:overview}

In this section and Sect.~\ref{s:lemma} we prove the following bound,
independently of Sect.~\ref{s:improved-complete}.

\begin{theorem} \label{t:simple}
The multi-rate primal-dual algorithm with the rate system of
Sect.~\ref{s:rs} and $r=(\sqrt{13}-1)/3$ achieves approximation ratio $\sqrt{13}+1$.
\end{theorem}

Let
\[
x=\sqrt{13}+1,\qquad r=(\sqrt{13}-1)/3,\qquad y=1/2.
\]
We use $xr=4$, $x=2+3r$, $x-y\ge4$, $x+y\ge5$, and $xr+2y\ge5$.
By Corollary~\ref{c:2C} and Lemmas~\ref{l:res}, \ref{l:arb}, and \ref{l:hits},
it suffices to prove:

\begin{lemma} \label{l:ratio}
Let $I$ be an inclusion-minimal cover of a strongly pliable family $\FF$.
If $\BB,\RR$ and $\MM$ are as in Lemma~\ref{l:hits} for the family $\CC$ of $\FF$-cores then
\begin{equation} \label{e:ratio}
\sum_{C\in\BB} d_I(C)+r\sum_{C\in\RR} d_I(C)
\le x\bigl(|\BB|+r|\RR|\bigr).
\end{equation}
\end{lemma}
Fix the hitting forest $\HH$ of $\FF$, the partition $\BB,\RR$, the matching $\MM$, and the corresponding rates. 
Let $K$ be the set of edges in $I$ that cover no core,
let $I'=I \sem K$, and let $\FF'=\FF^K$. Note that:
\begin{itemize}
\item
 $\FF$ and $\FF'$ have the same core family $\CC$ and 
$d_I(C)=d_{I'}(C)$ for all $C \in \CC$,
since $K$ covers no core.
\item
$I'$ is an inclusion-minimal cover of $\FF'$ (since $I$ is an inclusion-minimal cover of $\FF$) and 
$\FF'$ is strongly pliable (by Lemma~\ref{l:res}).
\item
If $A \in \FF'$ then $A \in \FF$, and if $A \in \FF'(C)$ then $A \in \FF(C)$ for any core $C$. 
Thus if a core $C_i$ hits a core $C_j$ in $\FF'$  then $C_i$ also hits $C_j$ in $\FF$.
\end{itemize}

It suffices to prove (\ref{e:ratio}) for $I'$ while retaining the original
$\HH$, $\BB,\RR$, $\MM$, and rates. We henceforth rename $\FF',I'$ as $\FF,I$.
Throughout the proof, ``hits'' refers to the fixed original forest $\HH$;
a hitting relation certified in the reduced family is also valid in $\HH$.

For every edge $e \in I$ we define two quantities:
\begin{itemize}
\item
The {\bf frequency} $f(e)$ of $e$ is the number of cores $e$ covers;
since the cores are pairwise-disjoint and every edge covers a core,
$f(e) \in \{1,2\}$.
\item
The {\bf weight} of $e$ is $w(e)=f_\BB(e)+rf_\RR(e)$, where
$f_\BB(e),f_\RR(e)$ count the blue and red cores covered by $e$.
Note that $w(e)\le f(e)$.
\end{itemize}

\noindent
{\bf Initial token assignment.} \\
{\em Assign $x$ tokens to every blue core and $rx$ tokens to every red core.}

\medskip

We will redistribute these tokens such that 
every $e \in I$ gets $w(e)$ tokens.
Since $w(I)$ equals the l.h.s. of (\ref{e:ratio}) 
and the total amount of tokens equals the r.h.s. of (\ref{e:ratio}), 
this will prove Lemma~\ref{l:ratio}.

The laminar witness family $\LL$ as in Lemma~\ref{l:lami} can be represented by 
a rooted tree $\TT$ on $\LL \cup \{V\}$, where $B$ is a child of $A$ if $A$ is the smallest 
set in $\LL \cup \{V\}$ that properly contains $B$.

\begin{definition}
An inclusion-minimal set in $\TT$ is a {\bf leaf}.
An {\bf internal} node is neither a root nor a leaf.
A set $A\in\TT$ is {\bf core-crossed} if some core crosses $A$.
A set {\bf owns a core $C$} if it is the inclusion-minimal set in $\TT$ that contains $C$,
and it is a {\bf terminal} if it owns a core or is the root $V$ of $\TT$. 
Let $D_A$ be the set of nodes in $A$ that belong to no child of $A$.
For a non-root set $A \in \TT$ let $e_A$ be the edge in $I$ that covers $A$
($e_A$ joins $D_A$ to the analogous region of its parent).
\end{definition}

\begin{lemma} \label{l:leaf-core}
Every leaf $A$ of $\TT$ contains  exactly one core that is a witness set for $e_A$. 
Thus we may assume w.l.o.g. that every leaf of $\TT$ is a core.
\end{lemma}
\begin{proof}
Every set in $\TT$ contains a core. 
Thus $A$ contains at least one core.
Suppose to the contrary that $A$ contains two distinct cores $C_1,C_2$.
Choose any edge $e_i\in I$ covering $C_i$, $i=1,2$.
One can see that the witness set of each $e_i$ must be $A$, namely $e_1=e_A=e_2$.
Since the cores are pairwise disjoint, we arrive at a contradiction. 
\end{proof}

\begin{definition} \label{d:special}
A leaf $C$ of $\TT$ is a {\bf safe leaf} if 
$\TT$ contains a halo-set $A$ of $C$ such that $A$ is core-crossed.
\end{definition}

We have the following property of safe leaves. 

\begin{lemma} \label{l:leaf-hit}
No leaf hits a safe leaf.
\end{lemma}
\begin{proof}
By the Sparse Crossing Property every safe leaf is hit by a unique core --
the core $C'$ that crosses its halo set that is in $\TT$.
But this core $C'$ is not a leaf of $\TT$. 
\end{proof}

\noindent
{\bf Moving tokens to safe leaves.}
\begin{enumerate}[(i)]
\item
If an edge of $\MM$ has a safe leaf as an endnode, this leaf gets $y$ tokens from the other endnode.
\item
Every safe red leaf gets $y$ tokens from each child in $\BB_0$.
\end{enumerate}
By Lemma~\ref{l:leaf-hit}, at most one endnode of an edge of $\MM$ is a safe leaf.
Every blue core gives at most $y$ tokens: a matched core only in (i),
and an unmatched core only in (ii), to its unique parent.
A safe core in $\BB_0$ gives no tokens, since its red parent is not a leaf.
Thus blue cores retain at least $x-y\ge4$ tokens;
safe matched blue leaves have $x+y\ge5$, safe red leaves have $rx+2y\ge5$,
and safe leaves in $\BB_0$ retain $x$.
Discarding surplus tokens gives:

\begin{corollary} \label{c:tokens}
The amount of tokens a core $C$ has is:
\begin{itemize}
\item
$5$ if $C$ is a safe red leaf or a safe matched blue leaf.
\item
$x$ if $C$ is a safe leaf in $\BB_0$; its parent in $\HH$ is red.
\item
$4$ otherwise.
\end{itemize}
\end{corollary}

\begin{definition}
A rooted subtree $\TT'$ of $\TT$ is {\bf hollow} if 
every terminal of $\TT'$ is a root or a leaf of $\TT'$.
For a terminal $S$ let $\TT_S$ denote the 
{\bf maximal hollow subtree} of $\TT$ rooted at $S$. 
\end{definition}

The next phase of our token distribution is by induction.
We assume $S\ne V$ when discussing the reserve at $S$;
the case $S=V$ is simpler, as no reserve is required. 
The maximal hollow subtrees form a hierarchy:
$\TT_{S'}$ is a child of $\TT_S$ if $S'$ is a leaf in $\TT_S$ (that is not a leaf of $\TT$).
Process this hierarchy bottom-up.
Within $\TT_S$, use the {\bf tokens of $\TT_S$} -- 
the tokens belonging to the leaves and root of $\TT_S$, 
to pay for its edges and leave a reserve at $S$. 
Then remove the processed descendants and continue upward.

Original leaf cores keep their red or blue labels. 
A processed non-leaf terminal becomes an artificial leaf of one of two types:
\begin{itemize}
\item
{\bf orange leaf}: has $5$ tokens; not an original leaf and its set is not core-crossed.
\item
{\bf yellow leaf}: \ has $4$ tokens; not an original leaf and its set is core-crossed.
\end{itemize}

Original leaf cores keep the tokens in Corollary~\ref{c:tokens}.
To be able to apply induction we will prove the following.

\begin{lemma} \label{l:main}
For a terminal $S$ that is not a leaf of $\TT$, 
the tokens of $\TT_S$ can be redistributed such that 
every edge in $\TT_S$ gets $w(e)$ tokens and $S$ retains:
\begin{itemize}
\item
$5$ tokens if $S$ is not core-crossed (then $S$ becomes an orange leaf).
\item
$4$ tokens if $S$ is core-crossed (then $S$ becomes a yellow leaf).
\end{itemize}
\end{lemma}

In the rest of this section we prove Lemma~\ref{l:main}. 
Let $U=\cup_{C \in \CC}C$ be the union of all cores. 
We will use the following lemma.

\begin{lemma} \label{l:A}
Let $A$ be a non-terminal set that is not core-crossed.
Then $D_A \cap U = \empt$. 
Consequently, $f(e_A)=1$, and every child $B$ of $A$ 
is a terminal that is not core-crossed with $f(e_B)=1$. 
\end{lemma}
\begin{proof}
No node in $D_A$ belongs to a core $C$ since then 
$A$ must own or cross $C$ (by Lemma~\ref{l:disjoint}),
contradicting that $A$ is a non-terminal set that is not core-crossed. 
Let $B$ be a child of $A$. 
Each of $e_A,e_B$ has exactly one end in $D_A$, thus $f(e_A)=f(e_B)=1$. 
Let $v$ be the end of $e_B$ in $D_B$. Then $v \in C$ for some core $C$.
No core can cross $B$, as then it would also cross $A$. 
Thus $B$ owns $C$, concluding the proof. 
\end{proof}

For illustration of Lemma~\ref{l:A} consider the case when no set in $\TT$ is core-crossed. 
Then we can easily prove the known bound 
$\sum_{C \in \CC} d_I(C)  =\sum_{e \in I} f(e) \leq  2 |\CC|$.
Assign $2$ tokens to each core.
Every leaf-to-root path in a hollow subtree has total frequency at most $2$,
so its leaf can pay for that path. 
The root terminal retains the tokens of a core it owns to continue the induction.


Fix a hollow subtree $\TT_S$, possibly already containing orange and yellow leaves,
and let $T_s$ denote its ordinary tree representation. 
A node $v$ of $T_s$ represents the set $S_v$. 
We will now modify $T_s$ as follows.
\begin{enumerate}
\item
If there is a node $v \ne s$ such that $S_v$ is not core-crossed and $v$ has $p \ge 2$ children,
then we drop all children and make $v$ an orange leaf with $5$ tokens.
By Lemma~\ref{l:A}, every child of $v$ is a leaf (thus has at least $4$ tokens) 
connected to $v$ by an edge of frequency $1$.
We pay for these edges, and give $p(4-1) =3p > 5$ tokens to $v$. 
After that every node $v \ne s$ of $T_s$ such that $S_v$ is not core-crossed 
is either a leaf or has a unique child that is a leaf.
\item
We shortcut every maximal path of nodes that have a unique parent and a unique child
(the sets of the shortcut nodes are also removed from $\TT$),
adjusting frequencies and weights accordingly.
The frequency and the weight of an edge that shortcuts a path are 
the sum of the frequencies and the weights of the edges in the path, respectively.
After that every node $v \ne s$ of $T_s$ such that $S_v$ is not core-crossed is a leaf.
\end{enumerate}

We continue to denote the resulting tree by $T_s$, that now has the following property.

\begin{corollary} \label{c:internal}
Any internal node $v$ of $T_s$ has at least two children and $S_v$ is core-crossed.
\end{corollary}

Note that $T_s$ has leaf-tokens in $\{5,x,4\}$,
and that only terminals (leaves and $s$) have tokens. 
Let $\tau(u)$ denote the amount of tokens of a terminal $u$. 
Let $L$ be the set of leaves of $T_s$. 
Let $q(S)$ be the number of cores $S$ owns or crosses.
Note that if $q(S) \ge 2$ then $S$ owns two cores or is core-crossed.
We say that $u$ is a {\bf good child of} $v$ if $S_u$ is not core-crossed and $f(uv) \le 2$.
A good child is a leaf and thus has at least $4$ tokens.
Thus we have:

\begin{corollary} \label{c:x-y}
Every good child $u$ of $v$ is a leaf that satisfies $\tau(u)-f(uv) \ge 2$.
\end{corollary}

In the next Sect.~\ref{s:lemma} we will prove the following lemma.
It says that ``heavy edges have other good properties''. 
A ``good property'' can be a terminal endnode with many tokens, 
or the existence of a good child that by Corollary~\ref{c:x-y} has a surplus of $2$ tokens.

\begin{lemma} \label{l:root}
Let $u$ be a child of $v$. Then $f(uv) \le 5$.
If $f(uv)=5$ then $u \in L$, $v=s$, $\tau(u) \ge w(uv)$, and $q(S) \ge 2$. 
Else, the following holds.
\begin{enumerate}[(i)]
\item
If $f(uv)=4$ then $u\in L$ and $\tau(u)\ge w(uv)$.
Moreover, $\tau(u)\ge w(uv)+1$ or $v=s$ and $q(S)\ge2$.
Furthermore, if $v \ne s$ then $v$ has a good child.
\item
If $f(uv)=3$ then $u \in L$ or $v=s$ and $q(S) \ge 2$. 
Furthermore, if $v \ne s$ then  $S_u$ is not core-crossed (so $\tau(u) \ge x$) 
or $v$ has a good child.
\item
If $f(uv)=2$ and $v \ne s$, then $u \in L$ or $v$ has a good child.
\end{enumerate}
\end{lemma}

We now use this lemma to finish the proof of Lemma~\ref{l:main}.

\begin{lemma} \label{l:Tv}
Let $v \ne s$ be a node of $T_s$. 
A leaf $v$ has tokens as specified in Corollary~\ref{c:tokens} if $v$ is an original leaf, 
$5$ if $v$ is an orange leaf, and $4$ if $v$ is a yellow leaf. 
If $v$ is an internal node of $T_s$ then  
the tokens of leaf descendants of $v$  can be redistributed such that 
every descendant edge of $v$ gets $w(e)$ tokens and $v$ gets $3$ tokens.
\end{lemma}
\begin{proof}
Note that for $v \ne s$ Lemma~\ref{l:root} implies that for any child $u$ of $v$:
\begin{enumerate}[(i)]
\item
If $f(uv)=4$ then $u \in L$, $\tau(u) \ge w(uv) +1$, and $v$ has a good child.
\item
If $f(uv)=3$ then $u \in L$ and: $\tau(u) \ge x$ or $v$ has a good child.
\item
If $f(uv)=2$ then $u \in L$ or $v$ has a good child.
\end{enumerate}
Use induction on subtree height. 
A processed internal child has $3$ tokens after paying for its descendant edges; 
a leaf has its stated token amount.
By Lemma~\ref{l:root} and since $v \ne s$, $f(uv) \le 4$.
Let $u$ be a child of $v$ such that $f(uv)$ is maximal. 
Consider the cases $f(uv) \in \{1, \ldots,4\}$.
\begin{itemize}
\item
{\boldmath $f(uv)=4$:}
By (i), $\tau(u) \ge w(uv) +1$, and $v$ has a good child $u'$.
Then  $v$ can get from $u,u'$ $1+2=3$ tokens.
\item
{\boldmath $f(uv)=3$:} 
By (ii), $u \in L$. 
If $v$ has a good child $u'$,
then $v$ can get from $u,u'$ $(4-3)+2=3$ tokens.
Assume that $v$ has no good child. 
Then $\tau(u) \ge x >4.5$ by (ii),
so $v$ can get $1.5$ tokens from $u$. 
Let $u'$ be another child of $v$. 
\begin{itemize}
\item
If $f(u'v)=3$ then $v$ can get $x-3> 1.5$ tokens from $u'$. 
\item
If $f(u'v)=2$ then $u' \in L$ by (iii), 
so $v$ can get $4-2=2$ tokens from $u'$.
\item
If $f(u'v) = 1$ then $v$ can get 
$3-1=2$ tokens from $u'$, by the induction hypothesis.
\end{itemize}
In all cases $v$ can get at least $1.5$ tokens from $u'$, hence at least $3$ tokens from $u,u'$. 
\item
{\boldmath $f(uv)=2$:}
By (iii), $u \in L$ or $v$ has a good child.
Then $v$ can get 
$4-2=2$ tokens from $u$ or from its good child and 
$3-2=1$ token from some other child, by the induction hypothesis.  
\item
{\boldmath $f(uv)=1$:}
By the induction hypothesis $v$ can get $3-1=2$ tokens from each child. 
Since $v$ has at least two children, it can get $4$ tokens. 
\end{itemize}
In every case the edge payments are valid since $w(e) \le f(e)$, 
except where the sharper weighted bound is used in the $f(uv)=4$ case.
Additional children can pay their incident edges without a deficit, using the same bounds. 
Hence at least three tokens remain at $v$. 
\end{proof}

From Lemmas~\ref{l:root} and~\ref{l:Tv} we get the following.

\begin{corollary} \label{c:z}
Let $z$ be a child of $s$.
The tokens in the subtree rooted at $z$ can pay for its descendant edges 
and the edge $zs$; if $q(S)=1$, one additional token can be given to $S$.
\end{corollary}
\begin{proof}
If $f(zs) = 5$, Lemma~\ref{l:root} guarantees the payment and implies $q(S) \ge 2$.
If $f(zs) \le 2$ Lemma~\ref{l:Tv} gives $3$ to an internal node $z$, 
while a leaf has at least $4$; at least $1$ token remains after paying for $zs$.
If $f(zs) =3$ and $z$ is internal, Lemma~\ref{l:root}(ii) gives $q(S)\ge2$, 
and the three-token reserve from Lemma~\ref{l:Tv} pays $zs$. 
Otherwise, $f(sz) \in \{3,4\}$ and $z$ is a leaf. 
When $q(S)=1$, a frequency-three leaf has at least four tokens, 
and a frequency-four leaf satisfies $\tau(z) \ge w(zs)+1$ by Lemma~\ref{l:root}(i).
\end{proof}

So far we did not use the tokens of cores owned by $S$.
If $S$ is core-crossed, an owned core supplies the $4$ tokens for a yellow leaf.
Otherwise, either $S$ owns at least two cores, or $q(S)=1$ and a child supplies
one extra token by Corollary~\ref{c:z}; this gives the $5$ tokens for an orange leaf.
This proves Lemma~\ref{l:main} and Theorem~\ref{t:simple}, subject to
Lemma~\ref{l:root}, proved in the next section.

\medskip

\noindent
{\bf Remark.}
To prove approximation ratio $5$ we do not need the forest structure of cores,
and from Lemma~\ref{l:root} we only need that
$f(uv) \le 5$, if $f(uv)=5$ then $u \in L$ and $v=s$, and:
\begin{enumerate}[(i)]
\item
If $f(uv)=4$ then $u \in L$ and if $v \ne s$ then $v$ has a good child.
\item
If $f(uv)=3$ and $v \ne s$ then $u \in L$.
\end{enumerate}
Assign $5$ tokens to every core. One can show by induction that for internal $v$
the tokens of leaf descendants of $v$ can be redistributed such that every descendant
edge of $v$ gets $f(e)$ tokens and $v$ gets $4$ tokens.
To see this, fix a child $u$ of $v$ such that $f(uv)$ is maximal. 
If $f(uv)=4$ then $v$ gets $1$ token from $u$ and $3$ tokens from a good child.
If $f(uv) \le 3$ then it is not hard to see that $v$ can get $2$ tokens from each child.  
This suffices to pay for all edges of $T_s$, 
and $s$ gets $5$ tokens from a core $S$ owns.

\section{Proof of Lemma~\ref{l:root}} \label{s:lemma}

We need a few simple lemmas for the proof of Lemma~\ref{l:root}.

\begin{lemma} \label{l:f=2}
Let $A \in \TT$.
If $f(e_A)=2$ then $A$ is a terminal or the parent of $A$ is a terminal. 
\end{lemma}
\begin{proof}
The ends of $e_A$ lie in distinct cores.
If neither core is owned by $A$ or its parent, 
both cores cross the parent, contradicting the Sparse Crossing Property.
\end{proof}

We will use the following consequence of Property $\ga^*$.

\begin{lemma} \label{l:D}
Let $A \in \TT$ be a non-terminal set that crosses a core $C$ 
and let $\{B_1, \ldots,B_p\}$ be the set of children of $A$ that cross $C$.
If the set $D=A \sem (B_1 \cup \cdots \cup B_p \cup C)$ is non-empty then
$A$ has a child $B \subs D$ that is not core-crossed. 
\end{lemma}
\begin{proof}
By Property $\ga^*$ (Lemma~\ref{l:ga*}), $D \in \FF$.
Thus $D$ contains a core $C_0$ distinct from $C$. 
Since $A$ is non-terminal, it does not own $C_0$, 
so some child $B$ of $A$ contains $C_0$.
As $C_0 \subs D$, $B \notin \{B_1, \ldots,B_p\}$.
Thus $B$ does not cross $C$. 
Since $\empt \ne C\sem A \subs C \sem B$, 
Lemma~\ref{l:disjoint} implies that $B\cap C=\empt$. 
Moreover, children of $A$ are pairwise disjoint, 
so $B$ is disjoint from every member of $\BB$. 
Consequently, $B\subs D$.
If $B$ crosses some core $C'$ distinct from $C$ then 
both $C,C'$ cross $A$ (since $A$ is non-terminal), 
contradicting the Sparse Crossing Property.
\end{proof}

\begin{figure} \centering \includegraphics[scale=0.44]{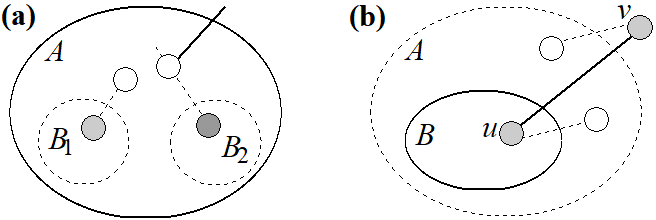}
\caption{Eliminating a non-leaf set $A$ that is not core-crossed.
Gray nodes are in $U$ and white in $V \sem U$. 
Sets and edges shown by dashed lines are removed.
In (b), $u,v$ must belong to distinct cores. 
}
\label{f:shortcuts} \end{figure}

Focus on some tree $\TT_S$. 
Some of the leaves of $\TT_S$ are original leaves of $\TT$ that are colored red or blue.
Any other leaf of $\TT_S$ is a non-leaf terminal set of $\TT$, 
that is colored either orange if it crosses no core and yellow otherwise. 
The following simple reduction will enable us to reduce the case analysis by assuming that 
any set $A \in \TT_S \sem \{S\}$ that is not core-crossed is a leaf. Suppose that $A$ is not a leaf.
By Lemma~\ref{l:A}, all children of $A$ are leaves and $D_A\cap U=\empt$.
\begin{enumerate}
\item
If $A$ has at least two children, then we drop all children and make $A$ an orange leaf, 
see Fig.~\ref{f:shortcuts}(a).
Note that this is exactly what we did at step $1$ when $T_s$ was modified.
\item
If $A$ has a unique child then we ``shortcut'' $A$ as follows:
we remove $A$ and replace $e_A,e_B$ by a new edge $uv$, where 
$u$ is the end of $e_B$ in $B$ and $v$ is the end of $e_A$ that is not in $A$, 
see Fig.~\ref{f:shortcuts}(b).
Note that $u,v$ cannot belong to the same core, as otherwise, this core crosses $A$.
Also note that $B$ remains a leaf of $\TT_S$ that now is a witness set for the new edge $uv$,
and that this operation does not change the total weight.
\end{enumerate}

Hence from now on we will assume that only leaves and the root of $\TT_S$ may not be core-crossed.
Namely, every internal (non-terminal) set of $\TT_S$ 
is core-crossed; a terminal set may or may not be core-crossed.

\begin{definition} \label{d:chain}
A sequence $\SS=(S_0, S_1, \ldots, S_\ell)$ of sets in $\TT_S$ is a (maximal) {\bf chain} if:
\begin{itemize}
\item
$S_1$ exists and $S_1,\ldots, S_\ell$ are non-terminals.
\item
$S_{i-1}$ is the unique child of $S_i$ for $i=1, \ldots,\ell$.
\item
Each of $S_0,S_{\ell+}$ is a terminal or has at least two children,
where $S_{\ell+}$ is the parent of $S_{\ell}$.
\end{itemize} 
The edge set of $\SS$ is $I_{\SS}=\{a_0b_1, \ldots,a_\ell b_{\ell+1}\}$, 
where $a_ib_{i+1}$ is the unique edge in $I$ that covers $S_i$ 
and $a_i,b_i \in S_i$ (possibly $a_i=b_i$). 
The {\bf weight} and the {\bf frequency} of a chain $\SS$ are 
$w(\SS)=w(I_\SS)$ and $f(\SS)=f(I_\SS)$, respectively.
\end{definition}

Recall that a leaf $C$ of $\TT$ is safe if $\TT$ contains a halo-set $A$ of $C$ such that $A$ is core-crossed.
If $C$ is the bottom set $S_0$ of some chain $\SS$ in $\TT_S$, then $A=S_1$ has the required property.
Thus we have the following.

\begin{corollary} \label{c:safe}
Let $C$ be an original leaf of $\TT$ 
that is the bottom set $S_0$ of some chain in $\TT_S$. 
Then $C$ is a safe leaf. 
\end{corollary}

\begin{lemma} \label{l:Si}
If a core $C$ crosses a set $S_i$ of a chain $\SS$ of length $\ell > i$
then $S_{i+1} \sem S_i \subs C$ and $\ell=i+1 \le 2$.
\end{lemma}
\begin{proof}
Note that $C$ crosses $S_i, \ldots,S_\ell$ since they all intersect $C$ but none of them owns $C$ 
(none of $S_{i+1}, \ldots, S_\ell$ is a terminal and $S_i$ crosses $C$). 
Since $i< \ell$, $C$ crosses $S_{i+1}$. 
We have $D=S_{i+1} \sem (S_i \cup C)=\empt$, 
as otherwise by Lemma~\ref{l:D} $S_{i+1}$ has at least two children.
If $\ell \ge i+2$ then we get the contradiction that $e_{S_{i+1}}$ goes 
from $S_{i+1} \sem S_i \subs C$ to $S_{i+2} \sem S_{i+1} \subs C$.
Applying the same argument at the first non-terminal set $S_1$ excludes $\ell \ge 3$.
\end{proof}

\begin{figure} \centering \includegraphics[scale=0.42]{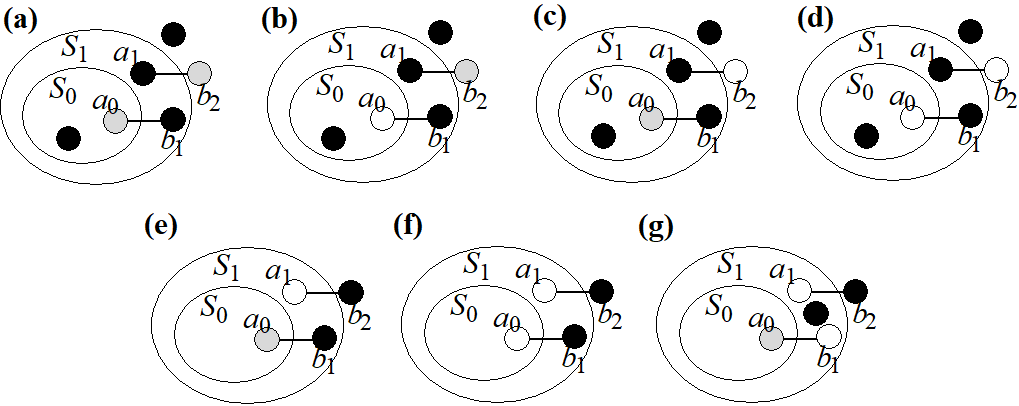}
\caption{Illustration to Lemma~\ref{l:C}.
Black nodes are in $C$, gray in $U \sem C$, and white in $V \sem U$.
}
\label{f:no-good} \end{figure}

\begin{lemma} \label{l:C}
Let $\SS$ be a chain of length $\ell \ge 1$ and let $C$ be a core.
\begin{enumerate}[(i)]
\item
If $C$ crosses $S_0$ then $\{b_1,a_1\} \subs S_1 \sem S_0 \subs C$
and $\SS$ is as in Fig.~\ref{f:no-good}(a,b,c,d). 
\item
If $C$ crosses $S_1$ and $b_{\ell+1} \in C$ then $S_0$ is not core-crossed 
and $\SS$ is as in Fig.~\ref{f:no-good}(e,f,g).
\end{enumerate}
\end{lemma}
\begin{proof}
We prove (i). 
We have $\ell=1$ and $\{b_1,a_1\} \subs S_1 \sem S_0 \subs C$ by Lemma~\ref{l:Si}. 
For each of $a_0,b_2$ the choice is whether it is in $U \sem C$ or not,
giving the four cases in the lemma. 

We prove (ii).
Since $b_{\ell+1} \in C$, $a_\ell \notin U$, 
and thus by Lemma~\ref{l:Si} $\ell=1$ and $S_0$ is not core-crossed. 
There are three such cases given in Fig.~\ref{f:no-good}(e,f,g),
determined by the possible choices
$b_1 \in C$ or $b_1 \notin U$ for $b_1$ and 
$a_0 \in U \sem C$ or $a_0 \notin U$ for $a_0$ ($a_0 \notin C$ since $S_0$ is not core-crossed), 
where the choice $a_0,b_1 \notin U$ is impossible. 
\end{proof}

\begin{figure} \centering \includegraphics[scale=0.46]{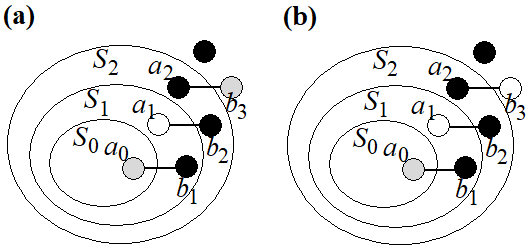}
\caption{$4,5$-chains in Lemma~\ref{l:5-chain}.
 Black, gray, white nodes have the same meanings as in Fig.~\ref{f:no-good}.}
\label{f:45} \end{figure}

\begin{lemma} [\cite{N-prel}] \label{l:5-chain}
Let $\SS$ be a chain. Then $f(\SS) \le 5$. Moreover:
\begin{enumerate}[(a)]
\item
If $f(\SS) =5$ then $\ell=2$, $S_0,S_{\ell+}$ are both terminals and
$\SS$ has the form as in Fig.~\ref{f:45}(a): 
$a_2, b_2, b_1$ belong to the core $C$ that crosses $S_1$,
$a_0, b_3$ belong to distinct cores that are also distinct from $C$, 
while other ends of the edges of the chain belong to no core.
\item
If $f(\SS) =4$ and one of $a_\ell,b_{\ell+1}$ is not in $U$
then $\ell=2$ and 
$\SS$ has the form as in Fig.~\ref{f:45}(b): 
$a_2, b_2, b_1$ belong to the same core $C$,
$a_0$ belongs to a core distinct from $C$, 
and $a_1,b_3$ belong to no core.
\end{enumerate}
\end{lemma}
\begin{proof}
If $\ell=1$ then $f(\SS) \le 4$ 
and if $f(\SS)=4$ then every edge of $\SS$ has both ends in $U$. 
Thus in both (a) and (b) we must have $\ell \ge 2$, and Lemma~\ref{l:Si} gives $\ell=2$. 
The middle edge cannot have both ends in $U$ (by Lemma~\ref{l:f=2})
and hence has frequency one.
The two outer edges each have frequency at most two, so $f(\SS) \le2+1+2=5$.
By Lemma~\ref{l:C}(i), if $S_0$ is core-crossed then $\ell=1$, 
thus let $C$ be the core that crosses $S_1$, 
and $S_0$ is not core-crossed. 
Note that $a_2,b_2 \in C$, by Lemma~\ref{l:Si}. 

We prove (a). 
Since $b_2 \in C$, $a_1 \notin C$, and by Lemma~\ref{l:f=2} $a_1 \notin U$. 
Since $f(\SS)=5$, $b_1 \in C$ and $a_0,b_3 \in U \sem C$,
arriving at the Fig.~\ref{f:45}(a) case.
By Lemma~\ref{l:f=2}, $S_0,S_{\ell+}$ are both terminals.

We prove (b).
Since $a_2 \in C$, the hypothesis of (b) gives $b_3 \notin U$. 
Thus we arrive at Fig.~\ref{f:45}(b) case.
\end{proof}

We say that a child $B$ of $A$ is a {\bf good child}
if $B$ is not core-crossed; note that $B$ is a leaf
and $f(e_B)\le2$.

\begin{corollary} \label{c:good-child}
Let $\SS$ be a chain such that $S_{\ell+} \ne S$ and let $C$ be the core that crosses $S_\ell$.
If $b_{\ell+1} \notin C$ then $S_{\ell+}$ has a good child.
\end{corollary}
\begin{proof}
Since $S_{\ell+} \ne S$, $C$ crosses both $S_\ell$ and $S_{\ell+}$. 
Let $D=S_{\ell+} \sem (B_1 \cup \cdots \cup B_p \cup C)$
where $B_1, \ldots, B_p$ are the children of $S_{\ell+}$ that cross $C$.
We have $b_{\ell+1} \in D$, thus by Lemma~\ref{l:D}
$S_{\ell+}$ has a good child contained in $D$. 
\end{proof}

In Lemma~\ref{l:root}, if $uv$ corresponds to a single edge after the
reductions of Fig.~\ref{f:shortcuts}, then $f(uv)\le2$.
If $f(uv)=2$ and $v\ne s$, then $u$ is a leaf: this is immediate in
Fig.~\ref{f:shortcuts}(b), and otherwise follows from Lemma~\ref{l:f=2},
since $v$ is non-terminal. Thus Lemma~\ref{l:root}(iii) holds in this case.
Excluding this case, we now restate Lemma~\ref{l:root} in terms of chains and prove it.

\begin{lemma} \label{l:root-chain}
Let $\SS$ be a chain. Then $f(\SS) \le 5$.
If $f(\SS)=5$ then $S_0$ is not core-crossed, $S_{\ell+}=S$, $\tau(S_0) \ge w(\SS)$, and $q(S) \ge 2$.
Else, the following holds.
\begin{enumerate}[(i)] 
\item
If $f(\SS)=4$ then $S_0$ is a leaf and $\tau(S_0)\ge w(\SS)$.
Moreover, $\tau(S_0)\ge w(\SS)+1$ or $S_{\ell+}=S$ and $q(S)\ge2$.
Furthermore, if $S_{\ell+} \ne S$ then $S_{\ell+}$ has a good child.
\item
If $f(\SS)=3$ then $S_0$ is a leaf or $S_{\ell+}=S$ and $q(S) \ge 2$.
Furthermore, if $S_{\ell+} \ne S$ then $S_0$ is not core-crossed
or $S_{\ell+}$ has a good child.
\item
If $f(\SS)=2$ and $S_{\ell+} \ne S$ then $S_0$ is a leaf or $S_{\ell+}$ has a good child.
\end{enumerate}
\end{lemma}
\begin{proof}
If $f(\SS)=5$ then $\SS$ is as in Fig.~\ref{f:45}(a).
We only prove that $\tau(S_0) \ge w(\SS)$, 
as other parts follow from Lemmas~\ref{l:5-chain}(a) and \ref{l:f=2}.
By Lemma~\ref{l:5-chain}(a), $b_1,b_2,a_2$ belong to the same core $C$ 
and $a_0$ belongs to a core $C_0$ distinct from $C$.
If $S_0$ is an orange leaf then $\tau(S_0)=5 =f(\SS) \ge w(\SS)$, and $S_0$ cannot be a yellow leaf.
If $S_0=C_0$ is an original leaf then $C_0$ is a safe leaf, by Corollary~\ref{c:safe}.
Thus if $C_0$ is red or matched blue then it has $5$ tokens.
Otherwise $C_0\in\BB_0$ has $x$ tokens, and its hitter $C$ is red,
so $w(\SS)\le2+3r=x$.
Thus in all cases $\tau(S_0)\ge w(\SS)$.

We prove (i), so $f(\SS)=4$. 
If $S_0$ is not core-crossed then it is a leaf.
If $S_0$ is core-crossed then by Lemma~\ref{l:C}(i) $\SS$ must be as in Fig.~\ref{f:no-good}(a), 
and then it is also a leaf, by Lemma~\ref{l:f=2}.
Thus $\tau(S_0)\ge4\ge w(\SS)$.
Outside Fig.~\ref{f:45}(b), Lemma~\ref{l:5-chain}(b) gives frequency $2$ at the top edge.
Lemma~\ref{l:f=2} forces $S_{\ell+}=S$, and its two endpoint cores give $q(S)\ge2$.
If $S_{\ell+}\ne S$, Corollary~\ref{c:good-child} gives a good child.
In Fig.~\ref{f:45}(b), by (hypothetically) coloring $b_3$ blue
and adding $1$ to the weight we obtain a $5$-chain $\SS'$ of weight $w(\SS')=w(\SS)+1$, 
that satisfies $\tau(S_0) \ge w(\SS')=w(\SS)+1$.

We prove (ii), so $f(\SS)=3$. 
If $S_0$ is not core-crossed then it is a leaf.
Else, by Lemma~\ref{l:C}(i) $\SS$ must be as in Fig.~\ref{f:no-good}(b,c), 
but in case (c) $S_0$ is also a leaf, by Lemma~\ref{l:f=2}.
Thus we are left with case (b), and then $S_{\ell+}=S$ and $q(S) \ge 2$, 
since $a_1,b_2$ belong to distinct cores. 
If $S_{\ell+} \ne S$ and $S_{\ell+}$ has no good child then 
by Lemma~\ref{l:C}(ii) and Corollary~\ref{c:good-child} 
the only possible case is as in Fig.~\ref{f:no-good}(e);
in this case $S_0$ is not core-crossed and thus is either an orange leaf or a safe original leaf, 
that has at least $x$ tokens.

We prove (iii), so $f(\SS)=2$. 
If $S_0$ is not core-crossed, then it is a leaf.
Else,  the only possible case is the one in Fig.~\ref{f:no-good}(d);
in this case $S_{\ell+}$ has a good child by Corollary~\ref{c:good-child}.
\end{proof}

This proves Lemmas~\ref{l:root}, \ref{l:main}, and Theorem~\ref{t:simple}.

\section{Improved approximation ratio} \label{s:improved-complete}

We keep the partition and matching of Lemma~\ref{l:hits}, but now let
\begin{equation} \label{e:improved-parameters}
x=\frac{3+\sqrt{37}}2,\qquad r=\frac{\sqrt{37}-1}6,\qquad y=1/2.
\end{equation}
Thus
\begin{equation} \label{e:improved-identities}
x=2+3r,\qquad xr=3+r,\qquad x-y\ge4,\qquad x+y\ge5,\qquad xr+2y=4+r.
\end{equation}
Also $r>1/2$ and $4+r>x>4.5$.
The main change is that a red core need only retain $3+r$ tokens, rather than $4$.
This lowers the reserve at some terminals, so we need the weighted versions
of the chain bounds below. The structural arguments of Sects.~\ref{s:overview},\ref{s:lemma}
are unchanged.

Make the reduction of Sect.~\ref{s:overview}, retaining the original
hitting forest, partition, matching, and rates. Assign $xr_C$ tokens to each
core and make the same transfers to safe leaves, with $y=1/2$.
By (\ref{e:improved-identities}), after discarding surplus tokens we have:
\begin{equation} \label{e:improved-core-tokens}
\tau(C)=\begin{cases}
5 & \mbox{if $C$ is a safe matched blue leaf},\\
4+r & \mbox{if $C$ is a safe red leaf},\\
x & \mbox{if $C$ is a safe leaf in $\BB_0$},\\
3+r & \mbox{if $C$ is red and is not a safe leaf},\\
4 & \mbox{otherwise}.
\end{cases}
\end{equation}
Every original safe leaf has at least $x$ tokens. A core owned by a non-leaf
terminal has $3+r$ tokens if it is red and $4$ otherwise.

\medskip
\noindent
{\bf The reserve invariant.}
For a set $A$ let $\chi(A)=1$ if $A$ is core-crossed and $0$ otherwise;
let $\eta(A)=1$ if $A$ owns exactly one core and this core is red, and $0$ otherwise.
Ownership is always in the original witness tree.
We will pay the edges of each maximal hollow subtree $\TT_S$ and leave at its root
\begin{equation} \label{e:reserve}
b(S)=5-\chi(S)-(1-r)\eta(S).
\end{equation}
This applies to non-leaf terminals; original leaf cores keep
(\ref{e:improved-core-tokens}). The reserve is $5$ or $4+r$ if $S$ is not
core-crossed, and $4$ or $3+r$ if $S$ is core-crossed. Only the case $\eta(S)=1$
has a smaller reserve than before.

Here $D_A$ refers to the original witness tree.
We need two endpoint observations for $\eta(B)=1$, where $R$ is the red core owned by $B$:
\begin{enumerate}[(i)]
\item
If $B$ is not core-crossed, the end of $e_B$ in $B$, if in $U$, belongs to $R$.
\item
If $B$ is core-crossed, its parent $A$ is non-terminal, and $f(e_B)=2$,
then the ends of $e_B$ are in $R$ and the core $C$ that crosses $B$.
Thus $w(e_B)\le1+r$.
\end{enumerate}
Indeed, a core meeting $D_B$ is owned by $B$ or crosses $B$, which proves (i).
In (ii), $C$ also crosses $A$, since otherwise $A$ would own $C$.
A core meeting $D_A$ crosses $A$, so the end in $D_A$ is in $C$ by sparse crossing.
The other end is in a distinct core meeting $D_B$, hence in $R$.

Apply the preliminary reductions of Sect.~\ref{s:lemma}.
When pruning a branching non-terminal that is not core-crossed,
each child can still contribute $3$ tokens: an original red leaf has at least
$3+r$ and pays weight $r$, while every other eligible leaf has at least $4$
and pays at most $1$. Hence $p\ge2$ children supply $3p\ge6$ tokens,
enough to leave $5$. The shortcut in Fig.~\ref{f:shortcuts}(b) preserves
the inner endpoint of its bottom leaf, so observation (i) remains valid;
its bottom is not core-crossed.
Shortcut the remaining chains to obtain $T_s$, as before.
Every good child still has surplus at least $2$ after paying its edge:
an original red leaf has at least $3+r$ and pays at most $1+r$;
every other good child has at least $4$ and pays at most $2$.

\medskip
\noindent
{\bf Weighted chain bounds.}
The token-independent statements of Lemma~\ref{l:root-chain} still hold.
We replace its token bounds by the following.

\begin{lemma} \label{l:improved-heavy}
Let $\SS$ be a chain.
\begin{enumerate}[(i)]
\item
If $f(\SS)=5$, then $\tau(S_0)\ge w(\SS)$.
\item
If $f(\SS)=4$, then $\tau(S_0)\ge w(\SS)$;
if $S_{\ell+}\ne S$ or $q(S)=1$, then $\tau(S_0)\ge w(\SS)+1$.
\item
If $f(\SS)=3$ and either $S_{\ell+}\ne S$ or $q(S)=1$, then $S_0$ is a leaf and
$\tau(S_0)\ge w(\SS)+1$.
If $S_{\ell+}\ne S$ has no good child, then also $\tau(S_0)\ge x$.
\end{enumerate}
\end{lemma}
\begin{proof}
For (i), consider Fig.~\ref{f:45}(a), where $S_0$ is not core-crossed.
An artificial bottom has $5$ tokens, or $4+r$ if $\eta(S_0)=1$;
in the latter case $a_0$ is in its red owned core by observation (i), so $w(\SS)\le4+r$.
An original bottom is safe. If red, it has $4+r$ tokens and $w(\SS)\le4+r$;
if matched blue, it has $5$ tokens.
Otherwise it is in $\BB_0$, has $x$ tokens, and is hit by the red middle core $C$,
so $w(\SS)\le2+3r=x$.

For (ii), unless $S_{\ell+}=S$ and $q(S)\ge2$ we have Fig.~\ref{f:45}(b).
Hypothetically coloring $b_3$ blue adds $1$ to the weight and gives the same
five-chain calculation, so $\tau(S_0)\ge w(\SS)+1$.
In the remaining cases $S_0$ is a leaf.
If it has at least $4$ tokens it pays $w(\SS)\le4$.
Otherwise it is artificial, core-crossed, and owns just one red core $R$;
an original bottom is safe and thus has at least $x$ tokens.
By Lemma~\ref{l:C}(i) we have Fig.~\ref{f:no-good}(a),
with weight $r_R+2r_C+r_D\le3+r$, where $D$ is the core of $b_2$.
Its $3+r$ tokens suffice.

For (iii), $S_0$ is a leaf by the structural proof of Lemma~\ref{l:root-chain}(ii).
Every such leaf has at least $4$ tokens, except an artificial core-crossed bottom
that owns just one red core $R$. In this case Fig.~\ref{f:no-good}(b) is excluded
by the hypothesis, leaving case (c), with weight $r_R+2r_C\le2+r$.
Thus its $3+r$ tokens leave a surplus of $1$.
If $S_{\ell+}\ne S$ has no good child, the same structural proof shows that
$S_0$ is not core-crossed. It is then an original safe leaf or an artificial
leaf with at least $4+r>x$ tokens.
\end{proof}

\medskip
\noindent
{\bf Redistribution in a hollow subtree.}
We retain the three-token reserve at every internal node of $T_s$.
Here are the changes to the proof of Lemma~\ref{l:Tv}, with $u$ a child of $v$
maximizing $f(uv)$.
For $f(uv)=4$, use the surplus $1$ of Lemma~\ref{l:improved-heavy}(ii)
and surplus $2$ of a good child.
For $f(uv)=3$, a good child again suffices by part (iii).
If there is no good child, $u$ has at least $x>4.5$ tokens, and another child
$u'$ contributes more than $1.5$: for frequency $3$ use the same bound;
for frequency $2$ it is a leaf and contributes at least $(3+r)-2>1.5$;
for frequency $1$ it contributes at least $3-1=2$.

For $f(uv)=2$, a good child supplies $2$ tokens, and any other child supplies $1$.
Suppose there is no good child. Then $u$ is a leaf.
Unless it is artificial, core-crossed, and has $\eta(S_u)=1$, it has at least $4$
tokens: an original red leaf would itself be a good child.
In the exceptional case $uv$ cannot represent a chain, since Fig.~\ref{f:no-good}(d)
would give a good child by Corollary~\ref{c:good-child}.
Nor is it the shortcut of Fig.~\ref{f:shortcuts}(b), whose bottom is not core-crossed.
Thus it is an original edge, and observation (ii) gives
\[
\tau(u)-w(uv)\ge(3+r)-(1+r)=2.
\]
Another child supplies at least $3-2=1$ token.
The case $f(uv)=1$ is unchanged. Additional children pay their edges without a deficit.
This proves the internal reserve of $3$ by the same induction as before.

Corollary~\ref{c:z} also remains valid.
An internal child of $s$ has $3$ tokens; every leaf has at least $3+r>3$.
These suffice for edges of frequency at most $3$, and leave $1$ token
when the frequency is at most $2$.
For frequencies $4,5$, and for the extra token when $q(S)=1$ and the frequency
is $3$ or $4$, use Lemma~\ref{l:improved-heavy}.
An internal child joined to $s$ by a frequency-three edge has $q(S)\ge2$,
as before.

It remains to verify (\ref{e:reserve}) using the unused tokens of cores owned by $S$.
If $S$ owns at least two cores, they supply $2(3+r)>5$ tokens.
Otherwise it owns one core, supplying $3+r$ if red and $4$ if blue.
These are exactly the required reserves when $S$ is core-crossed.
If $S$ is not core-crossed, $q(S)=1$, so a child supplies the extra token,
giving $4+r$ or $5$ as required.
This completes the induction on maximal hollow subtrees.
Every edge receives $w(e)$ tokens out of the initial total
$x\sum_{C\in\CC}r_C$, which proves Theorem~\ref{t:main} by Corollary~\ref{c:2C}.

\medskip
\noindent
{\bf Tightness.}
The constraints $x\ge2+3r$ and $xr\ge3+r$ imply
$(x-2)(x-1)\ge9$, hence $x\ge(3+\sqrt{37})/2$.
Thus the parameters above are optimal for this token assignment.
Both stated approximation guarantees are also tight for the corresponding rates.
Use the instance of \cite{N-tight}, with $p\ge2$ copies, renaming the axis node $r$ as $u$.
Its cores are $\{t_1\},\ldots,\{t_p\},\{u\}$ and
$C=\{z\}\cup\{x_i,y_i:1\le i\le p\}$.
The hitting forest is the star $C\rightarrow\{t_i\}$, $1\le i\le p$,
and the isolated node $\{u\}$.
Thus $C$ is red, all other cores are blue, and $\MM=\empt$.
For $\epsilon>0$ assign costs
\[
c(t_ix_i)=c(y_iu)=1+r,\qquad c(a_iy_i)=r,\qquad
c(t_ib)=1+\epsilon,\qquad c(uz)=1+r+\epsilon.
\]
At time $1$ precisely the links $t_ix_i,a_iy_i,y_iu$ become tight.
They form an inclusion-minimal cover, of cost $p(2+3r)$,
while $\{t_ib:1\le i\le p\}\cup\{uz\}$ is a cover of cost
$p(1+\epsilon)+1+r+\epsilon$.
The ratio of these costs tends to $2+3r$ as $p\rightarrow\infty$ and $\epsilon\rightarrow0$,
giving $\sqrt{13}+1$ and $(3+\sqrt{37})/2$ for the two choices of $r$.

\section{Concluding remarks}

We showed how the structure of strongly pliable families guides the choice of
dual growth rates, while hollow subtrees organize the analysis.
Natural questions are how far this approach can improve the ratio and whether
analogous structural choices of rates can improve primal-dual algorithms
for other network design problems.


\end{document}